\documentclass[copyright,creativecommons]{eptcs}
\providecommand{\event}{PDMC 2009} 
\providecommand{\volume}{??}
\providecommand{\anno}{20??}
\providecommand{\firstpage}{1}
\providecommand{\eid}{??}

\usepackage{breakurl}        
\usepackage{graphicx, subfigure}

\title{Formal Aspects of Grid Brokering\thanks{The research leading to these results has received funding from the 
European Community's Seventh Framework Programme FP7/2007-2013 under grant agreement 215483 (S-Cube).}}
\author{Attila Kert\'{e}sz
\institute{MTA SZTAKI \\ 1518 Budapest, P.O. Box 63, Hungary}
\email{attila.kertesz@sztaki.hu}
\and
Zsolt N\'{e}meth
\institute{MTA SZTAKI \\ 1518 Budapest, P.O. Box 63, Hungary}
\email{zsnemeth@sztaki.hu}
}
\def\titlerunning{Formal Aspects of Grid Brokering}
\def\authorrunning{A. Kert\'{e}sz \& Zs. N\'{e}meth}
\begin{document}
\maketitle

\begin{abstract}
Coordination in distributed environments, like Grids, involves selecting the most appropriate services, resources
or compositions to carry out the planned activities. Such functionalities appear at various levels of the infrastructure
and in various means forming a blurry domain, where it is hard to see how the participating components are related and what 
their relevant properties are. In this paper we focus on a subset of these problems: resource brokering in Grid 
middleware. This paper aims at establishing a semantical model for brokering and related activities by defining brokering 
agents at three levels of the Grid middleware for resource, host and broker selection. The main contribution of this 
paper is the definition and decomposition of different brokering components in Grids by providing a formal model using 
Abstract State Machines.
\end{abstract}


\section{Introduction}\label{intro}

Grid Computing has emerged from the area of Parallel and Distributed Computing more 
than a decade ago \cite{foster}. Since then numerous Grid projects have succeeded and Grids started 
to flourish. Recently several service Grid middleware solutions are used around the world \cite{egee,globus,unicore}, 
and a wide variety of resource management tools -- which are important components of grid computing infrastructure -- are available 
and under development \cite{rmstax}. Resource management in Grids has many aspects and involves different approaches and 
means. Among those we have investigated brokering and established a Grid resource brokering taxonomy \cite{brtax} to determine what 
properties brokers posses and what functionalities are desired for certain tasks. This survey shows that the currently available 
Grid resource management tools are built on different middleware components supporting different properties and named with a 
bunch of acronyms -- even the ones having similar purposes. This plethora of approaches formed the domain of  
Grid resource management into a grey box with blurry boundaries where neither the users nor the researchers can clearly
see how these tools are related and what their relevant properties are. Until the definitions and interrelations are clarified, 
further development and interoperability cannot be facilitated. Therefore, in an earlier work we aimed at an informal definition 
as Grid resource management anatomy in \cite{anatomy}. Present work can be considered as a continuation, investigating how the area 
of Grid resource management can be formalized and what essential layers, functionalities can be separated based on the formal model.

A former work this paper is built on presents a formal definition for Grid Computing \cite{gridasm}. That time there had been several definitions 
for Grid Computing without the ability of making a clear distinction between Grids and other distributed systems. The paper concluded that 
Grids cannot be defined purely by their properties rather, their runtime semantics make the real difference. Based on the analysis, a formal 
definition was given for Grid Computing revealing its essential and characteristic functionalities. The aim and methodology of this paper is similar: 
establishing a \emph{formal, semantic model} for Grid resource management using \emph{Abstract State Machine}. We \emph{extend} the formal model 
for Grids defined in \cite{gridasm} by \emph{classifying} brokering components into three categories and \emph{defining} three agents for resource 
management at different levels of Grid systems. 

We are not aware of any other works that investigate formal models specifically for grid resource manager components. 
Bratosin et al. proposed a reference model for Grid architectures based on colored Petri nets in \cite{gridrefmod}. Though they 
provide a definition for job scheduling, they do not detail brokering steps and mechanisms at different levels. Altenhofen 
et al. investigated Service Oriented Architectures in \cite{asmsoa}, more specifically service discovery, mediation and 
composition. These components have some similar functionalities but this work is more focused on a unified, higher 
level service framework, and do not explore resource manager components. B\"orger et al. proposed an ASM model for workflows in 
\cite{wfasm}. The work presents workflow interpretations and transitions, which are related to our model, but they stay at the 
application level and do not deal with brokering at job level whereas our model targets the middleware below the application level.

In the following section we give a brief introduction of the formal Abstract State Machine method, and in Section \ref{gridasm} we 
summarize the formal model for Grid Computing introduced in \cite{gridasm} and describe our modified model. 
In Section \ref{brokerasm} we present and describe the extensions for Grid brokering 
components and in Section \ref{refinement} we refine agents responsible for broker and host selection.
Finally Section \ref{conclusions} concludes the paper.

\vspace{5mm}

\section{Abstract State Machines}\label{asm}

ASM represents a mathematically well founded framework for system design and
analysis \cite{asm}. It is able not just to model a working mechanism precisely 
but also to reveal the highly abstract nature of a system, i.e. to grasp the semantics \cite{ea}. 
Furthermore -- unlike many other state based modeling methods –-, it can easily be tailored to the required level of abstraction. 
Logician's structures applied in ASMs offer an expressive, flexible and complete way of
state description. The basic sets and the functions interpreted on them can be
freely chosen to the required level of complexity and precision \cite{asm}. 

In ASM, a signature (or vocabulary) is a finite set of function names, each
of fixed arity. Furthermore, it also contains the symbols $true, false, undef, =$
and the usual Boolean operators. A state $A$ of signature $\Upsilon$ is a nonempty set
$X$ together with interpretations of function names in $\Upsilon$ on $X$. $X$ is called the
superuniverse of $A$. An $r$-ary function name is interpreted as a function from $X^r$
to $X$, a basic function of $A$. A $0$-ary function name is interpreted as an element
of $X$ \cite{asmweb}.
A location of $A$ (can be seen like the address of a memory cell) is a pair $l$ $=$ ($f$, $a$), 
where $f$ is a function name of arity $r$ in vocabulary $\Upsilon$ and $a$ is an $r$-tuple 
of elements of $X$. The element $f(a)$ is the content of location $l$.
An update is a pair $a$ $=$ ($l$, $b$), where $l$ is a location and $b$ is an element of
$X$. Firing $a$ at state $A$ means putting $b$ into the location $l$ while other locations
remain intact. The resulting state is the sequel of $A$. It means that the interpretation 
of a function $f$ at argument $a$ has been modified resulting in a new state. 
ASMs are defined as a set of rules. An update rule $f(a)$ $:=$ $b$ causes an update
(($f$, $a$), $b$), i.e. hence the interpretation of function $f$ on argument $a$ will result $b$.
It must be emphasized that both $a$ and $b$ are evaluated in $A$. 

The nullary $Self$ function allows an agent to identify itself among other
agents. It is interpreted differently by different agents (that is why it is not a
member of the vocabulary.) An agent a interprets $Self$ as $a$ while an other agent
cannot interpret it as $a$. The $Self$ function cannot be the subject of updates.
A conditional rule $R$ is of form

\newcommand{\keyw}[1]{{\bf #1}}
\begin{tabbing}
\quad \=\quad \=\quad \kill
\keyw{if} $c$  \\
\> \keyw{then} $R_1$ \\
\keyw{else} \\
\> $R_2$ \\
\keyw{endif} \\
\end{tabbing}

To fire $R$ the guard $c$ must be examined first and whenever it is true $R_1$ 
otherwise, $R_2$ must be fired. A block of rules is a rule and can be fired 
simultaneously if they are mutually consistent. Some applications may require 
additional space during their run therefore, the $reserve$ of a state is the 
(infinite) source where new elements can be imported from by the following construct

\begin{tabbing}
\quad \=\quad \=\quad \kill
\keyw{extend} $U$ \keyw{by} \( v_1, \ldots, v_n \) \keyw{with} \\
\> $R$ \\
\keyw{endextend} \\
\end{tabbing}

meaning that new elements are imported from the $reserve$ and they are
assigned to universe $U$ and then rule $R$ is fired \cite{asmweb}.

The basic sequential ASM model can be extended in various ways like non-deterministic 
sequential models with the choice construct, first-order guard expressions, 
one-agent parallel and multi-agent distributed models. A distributed ASM \cite{asm} 
consists of a finite set of single-agent programs \(  \Pi_{\mathrm{n}} \) called modules, 
a signature $\Upsilon$, which includes each Fun($\Pi_{\mathrm{n}}$)-\{$Self$\}, i.e. 
it contains all the function names of each module but not the nullary $Self$ function, and 
a collection of initial states.

As it can be seen, ASM states are represented as (modified) logician's structures, i.e. 
basic sets (universes) with functions interpreted on them. Structures are modified in ASM 
to enable state transitions for modeling dynamic systems. Applying a step of ASM $M$ to state 
(structure) $A$ will produce another state $A'$ on the same set of function names. If the 
function names and arities are fixed, the only way of transforming a structure is to change 
the value of some functions for some arguments. Therefore, the most general structure transformation 
(ASM rule) is a guarded destructive assignment to functions at given arguments \cite{asm}.

Refinement \cite{asm} is defined as a procedure where abstract and more
concrete ASMs are related according to the hierarchical system design. At
higher levels of abstraction implementation details have less importance whereas
they become dominant as the level of abstraction is lowered giving rise to practical 
issues. Its goal is to find a controlled transition among design levels.

\section{ASM for Grid Computing}\label{gridasm}

\begin{figure}
\centerline{
\subfigure[]{
\includegraphics[width=7.0cm]{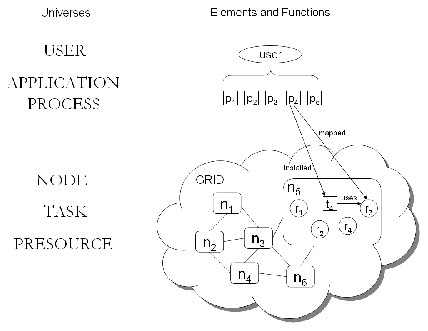}}
\subfigure[]{
\includegraphics[width=7.0cm]{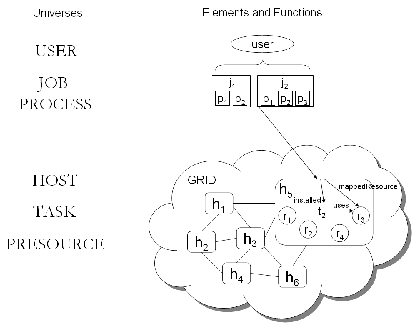}}
}
\caption{(a) Basic elements of the ASM model for Grids, and (b) the modified ASM model.}
\label{fig1}
\end{figure}

Before we define our model, we summarize the ASM for Grids defined in \cite{gridasm}. Figure \ref{fig1} (a) shows the important elements of 
this model. The ASM universes of the model are depicted on the left of the figures, and on the right a graphical representation 
of the connections of some elements of these universes and the most relevant functions governing process execution are shown. In this model 
user applications consist of one or more processes, while Grids consist of several nodes having one or more resources. During the 
execution of the user application first an agent maps the actual process of the application to a resource in the Grid, then the 
process is installed on the node of the resource as a task, which starts to use the resource. When all the processes of the application 
finished using their resources, the application is finished.

We extend this formal model by introducing Grid brokering at different levels. 
The basic model of grid systems introduced in \cite{gridasm} is presented in a slightly modified form here. The modification is 
indicated by introducing more practical issues related to realization; aligning the model to the terminology and naming conventions 
of grid brokering; and finally by experiences in Grid Computing since the paper was published. These modifications do not 
invalidate or alter the content and conclusion of the initial model just add more relevant details. The modifications are 
shown in Figure \ref{fig1} (b). In the following subsections we define the 
basic elements of our proposed extended formal model based on ASM: the universes, the signature and the rules.

\subsection{Universes and signature}

To define our formal framework, first we need to examine real service Grid systems. Certain objects of the physical reality are
modeled as elements of universes and relationships between real objects are represented as functions and relations. In Grid 
systems users (universe $USER$) define their applications in the form of jobs (universe $JOB$), which is the most 
typical computing paradigm for Grids hence, we restrict our model to this case. A job consists of one or more processes 
(universe $PROCESS$). The installed instances of processes are called as tasks (universe $TASK$), which can be run on different 
hosts (universe $HOST$). Hosts are the building blocks of Grid systems, and typically a job is sent to a host for execution. A host 
may have several nodes (e.g. when a host is a cluster), and nodes have certain resources that processes require to run. Since nodes 
are usually invisible (and unmanageable) for higher level tools, therefore we neglect them in our model. In this way one or more physical resources 
(universe $PRESOURCE$) belong to a host, which also determines the physical location (universe $LOCATION$) of the resources. 
The processes of jobs require some of these resources to run. Users should select a host according to these resource requirements, which we 
call as abstract resources (universe $ARESOURCE$). Information on the physical resources of the hosts can be gathered by querying the information 
system of a Grid. 

Once a job is \emph{submitted} to a host, it is mapped to physical resources during execution. While a resource is busy, the mapped process is 
in \emph{waiting} state. When the resource becomes free, the process starts using it and enters \emph{running} state. Process termination implies 
a \emph{done} state in case of successful run, and a \emph{failed} state in case of an error. In general, Grid authorization allows users to log 
in to some hosts and validates user privileges to use some resources of some hosts \cite{gridauth}. The requested (abstract) and the physical 
resources have certain attributes (universe $ATTR$). Compatibility between an abstract and a physical resource means the physical resource can 
satisfy the process requirement. According to this informal description, the following functions are used in the model: \\

\noindent job: \( PROCESS \rightarrow JOB \) \\
user, globaluser, localuser: \( JOB \rightarrow USER \) \\
submitted: \( JOB \times HOST \rightarrow \{true, false\} \) \\
procRequest: \( PROCESS \times ARESOURCE \rightarrow \{true, false\} \) \\
uses: \( PROCESS \times PRESOURCE \rightarrow \{true, false\} \) \\
mapped: \( PROCESS \rightarrow LOCATION \) \\
belongsTo: \( PRESOURCE \times HOST \rightarrow \{true, false\} \) \\
installed: \( TASK \times LOCATION \rightarrow \{true, false\} \) \\
attr: \( \{ ARESOURCE, PRESOURCE \} \rightarrow ATTR \) \\
location: \( PRESOURCE \rightarrow LOCATION \) \\
handler: \( PRESOURCE \rightarrow PROCESS \) \\
type: \( PRESOURCE \rightarrow ATTR \) \\
compatible: \( ATTR \times ATTR \rightarrow \{true, false\} \) \\
canLogin: \( USER \times HOST \rightarrow \{true, false\} \) \\
canUse: \( USER \times PRESOURCE \rightarrow \{true, false\} \) \\
jobState: \( JOB \rightarrow \{submitted, running, waiting, done, failed\} \) \\
procState: \( PROCESS \rightarrow \{running, waiting\} \) \\
event: \( TASK \rightarrow \{start, abort, terminate\} \) \\
mappedHost: \( JOB \rightarrow HOST \) \\
mappedResource: \( PROCESS \times ARESOURCE \rightarrow PRESOURCE \) \\

\subsection{Initial state}

We assume that $k$ processes belong to a job of a user. The job and its processes have some requirements, and 
no process and job is mapped to any resource or host. Therefore the states of the jobs and processes are undefined. 
In the following we define the initial state of our model: \\

\noindent \( \exists p_1, p_2, \ldots, p_k \in PROCESS: job(p_i) \neq undef, 1 \leq i \leq k \) \\
\( \forall p_i, 1 \leq i \leq k: \) user($p_i$) \( = u \in USER \) \\
\( \forall p_i, 1 \leq i \leq k: \exists ar \in ARESOURCE: \) procRequest($p_i, ar$) \( = true \) \\
\( \forall p_i, 1 \leq i \leq k: \exists pr \in PRESOURCE: \) uses($p_i, pr$) \( = false \) \\
\( \forall j: \) mappedHost($j$) \( = undef \) \\
\( \forall p_i , 1 \leq i \leq k: \) task($p_i$) \( = undef \) \\
\( \forall p_i , 1 \leq i \leq k: \) mapped($p_i$) \( = undef \) \\
\( \forall j: \) jobState($j$) \( = undef \) \\
\( \forall p_i , 1 \leq i \leq k: \) procState($p_i$) \( = undef \) \\
\( \forall u \in USER, \exists pr_1, pr_2, \ldots, pr_m  \in PRESOURCE: \) \\
canUse($u$, $pr_i$) \( = true, 1 \leq i \leq m \) \\

After we have defined the universes and the signature, in the following we give the rules of our model that constitute a module, 
i.e. a program that is executed by each agent in the model. The model presented here is a distributed multi-agent ASM where 
agents are jobs, i.e. elements from the $JOB$ universe. The working behavior of the brokering model is depicted from the 
perspective of the jobs hence, the self function is represented as $j$ and means the identity of a job, i.e. it can identify 
itself among other agents. It is interpreted differently by different agents.

\subsection{Rule 1: Resource selection}

According to Figure \ref{fig1} (b), when the job is sent to a host, the required resources need to be selected that 
are used by the processes of the job. During job execution, a task of each process is installed to 
the location of the required and selected resource. The precondition of resource selection is that the 
process of the job should be able to use the mapped resource. In case of the process can directly access 
the physical resource ($r_d$) the execution (resource usage) is automatically started, otherwise a local handler process 
should provide the execution platform (i.e. the additional software or service). If this handler process does not exist, 
it should be started before execution. The agent responsible for resource mapping needs to ensure that the chosen resource 
fulfills the abstract resource requirement of the process. Here is the formal definition: \\

\begin{tabbing}
\quad \=\quad \=\quad \=\quad \=\quad \=\quad \=\quad \kill
\keyw{let} $h$ $=$ mappedHost($j$) \\
\keyw{let} job($p$) $=$ $j$ \\
\keyw{let} $pr$ $=$ mappedResource($p$, $ar$) \\
\keyw{if} ( \( \exists ar \in ARESOURCE \) ): \\
\> procRequest($p$, $ar$) \( = true ~ \& ~ pr \neq undef ~ \& \) canUse(user($p$), $pr$) \( = true \) \\
\> \> \keyw{then} \keyw{if} type($pr$) $=$ $r_d$ \\
\> \> \> \keyw{then} mapped($p$) $:=$ location($pr$) \\
\> \> \> installed(task($p$), location($pr$)) \( := true \) \\
\> \> \keyw{else if} ( \( \neg \exists p' \in PROCESS \) ): handler($pr$) $=$ $p'$ \\
\> \> \> \> \keyw{extend} $PROCESS$ \keyw{by} $p'$ \\
\> \> \> \> \keyw{with} mapped($p'$) $:=$ location($pr$) \\
\> \> \> \> installed(task($p'$), location($pr$)) \( := true \) \\
\> \> \> \> handler($pr$) \( := p' \) \\
\> \> \> \> \keyw{do forall} \( ar \in ARESOURCE \) \\
\> \> \> \> \> procRequest($p'$, $ar$) \( := false \) \\
\> \> \> \> \keyw{enddo} \\
\> \> \> \keyw{endextend} \\
\> \> \keyw{endif} \\
\> \> procRequest(p, ar) := false \\
\> \> uses(p, pr) := true \\
\keyw{endif} \\
\end{tabbing}

\noindent \(  \Pi_{\mathrm{resource\_mapping}} \) \\

\begin{tabbing}
\quad \=\quad \=\quad \=\quad \kill
\keyw{if} ( \( \exists ar \in ARESOURCE, \exists p \in PROCESS, \exists h \in HOST \) ): \\
job($p$) $=$ $j$ $\&$ mappedResource($p$, $ar$) \( = undef ~ \& \) \\
procRequest($p$, $ar$) \( = true ~ \& \) $h$ $=$ mappedHost($j$) \\
\> \keyw{then} \keyw{choose} $pr$ \keyw{in} $PRESOURCE$ \\
\> \> \keyw{satisfying} compatible(attr($ar$), attr($pr$)) $\&$ belongsTo($pr$, $h$) \( = true \) \\
\> \> \> mappedresource($p$, $ar$) $:=$ $pr$  \\
\> \keyw{endchoose}  \\
\keyw{endif} \\
\end{tabbing}

Here, we note that though generally a job runs on a host (if it is a parallel job of communicating processes, it runs 
on a number of nodes of this host parallelly), some middleware tools may enable co-allocation of parallel processes on 
nodes of different hosts. We do not deal with this situation, since it is rarely used and supported, but further 
refinement of our model could represent such cases.

Before job execution it is necessary to authenticate users. In Service Grids users are authenticated by proxies of 
grid certificates \cite{gridauth}. A local process is responsible for validating these proxies by mapping global 
users to local ones having the same privileges. The related formalism of user mapping is similar to the one presented 
in \cite{gridasm}.

%

\subsection{Rule 2: State transition}

In this subsection we define, how job states are evolving during execution: \\

\begin{tabbing}
\quad \=\quad \=\quad \kill

\keyw{if} ( \( \exists h \in HOST \) ): submitted($j$, $h$) $= true$  \\
\> \keyw{then} jobState($j$) $:=$ $submitted$ \\
\keyw{endif} \\

\keyw{if} ( \( \exists p \in PROCESS \) ): job($p$) = $j$ $\&$ mapped($p$) $\neq$ $undef$  \\
\> \keyw{then} procState($p$) $:=$ $waiting$ \\
\> jobState($j$) $:=$ $waiting$ \\
\keyw{endif} \\

\keyw{if}  ( \( \exists pr \in PRESOURCE, \exists p \in PROCESS \) ): job($p$) = $j$ $\&$ uses($p$, $pr$) $= true$  \\
\> \keyw{then} procState($p$) $:=$ $running$ \\
\> jobState($j$) $:=$ $running$ \\
\keyw{endif} \\

\keyw{if}  ( \( \exists p \in PROCESS, \exists t \in TASK, \exists pr \in PRESOURCE, \exists h \in HOST \) ): \\
\> uses($p$, $pr$) \( = true ~ \& ~ belongsTo(pr, h) = true \) \\
\> $\&$ installed($t$, $h$) \( = true ~ \& \) event($t$) $=$ $abort$  \\
\> \> \keyw{then} jobState($j$) $:=$ $failed$ \\
\> \> $PROCESS$($p$) $:=$ $false$ \\
\keyw{endif} \\
\end{tabbing}

Though in general, process spawning could cause additional resource requests for job execution in a host, we do not detail this 
in our model, and keep it as abstract as possible, since at the level of grid brokering process communications and spawning are 
invisible. In order to handle these situations, we assume that resource requests of spawned processes are known a priori. State 
transitions related to job termination are formalized in Rule 3.

\subsection{Rule 3: Termination}

Job execution is terminated under the following conditions: \\

\begin{tabbing}
\quad \=\quad \=\quad \=\quad \kill
\keyw{if} ( \( \exists p \in PROCESS \) ): \\
\> job($p$) = $j$ $\&$ procState($p$) $=$ $running$ $\&$ event(task($p$)) $=$ $terminate$ \\
\> \> \keyw{then} $PROCESS$($p$) $:=$ $false$ \\
\keyw{endif} \\

\keyw{if} ( \( \exists p_1, \ldots, p_m \in PROCESS \) ): job($p_i$) $=$ $j$ $\&$ jobState($j$) $=$ $failed$, $1 \leq i \leq m$ \\
\> \keyw{then} $PROCESS$($p_i$) $:= false$ \\
\keyw{endif} \\

\keyw{if} ( \( \neg \exists p \in PROCESS \) ): job($p$) $=$ $j$ $\&$ jobState($j$) $=$ $running$ \\
\> \keyw{then} jobState($j$) $:= done$ \\
\keyw{endif} \\
\end{tabbing}

\section{ASM for Grid Brokering}\label{brokerasm}

This section focuses on middleware components responsible for brokering in Grids. In our ASM model these components are represented 
by agents (also called as modules). First we give an informal overview of these components and their roles in Grids and demonstrate 
their usage in a typical Grid application execution scenario. In the following subsections we show how these components can appear as 
agents in the formal model described above. Furthermore we emphasize how these brokering components contribute to Grid Interoperability, 
i.e. how they support transparent job submissions to different, separated Grids. 

At the lowest level of Grid resource management we can find local resource managers (also called as schedulers or cluster managers, 
e.g. PBS \cite{pbs}) that were taken from high-performance and distributed computing, and now generally used in 
Grid Systems. Their goal is to schedule and manage single and parallel programs and their processes locally. This local management 
is formalized in Rule 1 of or model. In this case interoperability is not supported at all. Without additional brokering components 
users need to choose from the available hosts manually relying on mostly static information.

One level above, a grid Resource Management System (RMS), also called as a Grid resource broker, is needed to automate host selection. 
It can be an internal middleware service, or an external tool that uses other middleware components or services. (Note that 
the word "resource" is used differently in our model as in the expression "resource broker". In our model we call the 
computing and storage elements of Grids as "hosts", and "resources" are the physical components of the hosts, e.g. processor 
and memory.) While local managers usually schedule user programs in a grid host (e.g. in a cluster) among free processors, 
the Grid broker schedules jobs at the level of Grid middleware by selecting a host that best matches 
the requirements of these jobs. (Thus, the selected host can also be a cluster with a local manager.) 
Therefore, a broker is also called as a meta-scheduler -- more information on broker naming conventions and their connections can be
found in \cite{anatomy}. Some of them support different middleware solutions, job types, agreements or various quality 
of service (QoS) attributes. Furthermore different brokers may be connected to different hosts and Grids. A taxonomy in \cite{brtax} introduces 
these properties and shows the differences among the currently used brokers, their properties, organization and connections and among their level of 
interoperability. In our future work we also plan to represent interoperability as a metric in our model in order to categorize and 
differentiate various brokering components.

With the help of grid brokers, host selection is automated, but users are still bound to separate grid islands (i.e. grid systems that are 
complete systems on their own but closed to any form of interoperability between each other, either by technology, compatibility, administrative 
or other restrictions) managed by their own brokers. Nevertheless users have the ability to select manually, which broker and Grid they would like 
to use (even static information on broker properties are available in form of manuals or taxonomies e.g. in \cite{brtax}). In order to achieve the 
highest level of interoperability broker selection should also be automated. Therefore at the highest level we can find meta-brokering \cite{mb}, 
which is a novel approach that introduces another layer above current grid brokers in order to facilitate inter-grid load balancing and interoperable 
brokering. The Grid Meta-broker sits on top of Grid brokers and uses meta-data to decide which broker should be selected for a user's job. To demonstrate the interoperation 
of these brokering components, we describe a typical Grid usage scenario for a job execution that requires the following steps:

\begin{enumerate}

\item The user defines its application as jobs, also stating the requirements of its execution.

\item The user requirements of the job is examined by the meta-broker, and mapped to the properties of the available 
brokers. A proper broker, that is able to submit the job, is selected for submission.

\item The selected broker examines the resource requirements of the job and matches them to the physical resources of 
the available hosts. A host having all the required resources is selected for execution. 

\item The agent on the selected host (the local resource manager) maps the resource requirements of the job to the 
available physical resources during execution.

\end{enumerate} 

In the following subsections we define two more rules to model the informal description and discussion above. We need 
additional universes and functions to incorporate brokering into our model.

\subsection{An additional rule for Grid Brokering}

Brokers (universe $BROKER$) are responsible for host selection, therefore hosts are managed by brokers, which can have 
different properties (universe $PRORERTY$) that users may require for job execution. A user should select a broker for 
its job according to these requirements (universe $REQUIREMENT$). Furthermore we place universe $ARESOURCE$ as a subset 
of universe $REQUIREMENT$, since the elements of both sets represent user requirements, and universe $PRESOURCE$ can be 
a subset of universe $PRORERTY$, because physical resources can be regarded as host properties. The following functions 
are added to the model: \\

\noindent request: \( JOB \times REQUIREMENT \rightarrow \{true, false\} \) \\
submitted: \( JOB \times \{ HOST, BROKER \} \rightarrow \{true, false\} \) \\
manages: \( HOST \times BROKER \rightarrow \{true, false\} \) \\
have: \( BROKER \times PRORERTY \rightarrow \{true, false\} \) \\
attr: \( \{ REQUIREMENT, PRORERTY \} \rightarrow ATTR \) \\

\noindent We extend the initial state by: \\

\noindent \( \forall j: \exists r \in REQUIREMENT: \) request($j, r$) \( = true \) \\

\noindent We extend Rule 2 with the following state changes: \\

\begin{tabbing}
\quad \=\quad \=\quad \kill
\keyw{if} ( \( \exists b \in BROKER \) ): submitted($j$, $b$) $= true$ \\
\> \keyw{then} jobState($j$) $:=$ $submitted$  \\
\keyw{endif} \\

\keyw{if} ( \( \exists h \in HOST \) ): submitted($j$, $h$) $= true$  \\
\> \keyw{then} jobState($j$) $:=$ $waiting$ \\
\keyw{endif} \\
\end{tabbing}

\begin{figure}
\centerline{
\subfigure[]{
\includegraphics[width=7.0cm]{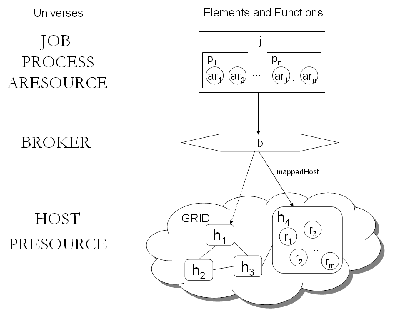}}
\subfigure[]{
\includegraphics[width=7.0cm]{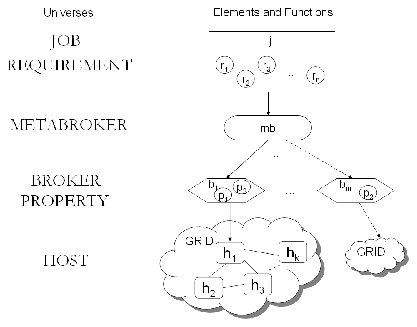}}
}
\caption{(a) Grid brokering and (b) Meta-brokering in the ASM model.}
\label{fig3}
\end{figure}

Once a broker is selected by the user, it should find an execution host. The precondition of this host selection process 
is that the user of the job should be able to use the required resources of the selected host. The broker agent responsible 
for host mapping needs to ensure that the chosen host has all the resources requested by the processes of the job. This 
additional component responsible for Grid brokering is highlighted in Figure \ref{fig3} (a). In the following we state the 
formal definition: \\

\noindent \emph{Rule 4: Host selection} \\

\begin{tabbing}
\quad \=\quad \=\quad \kill
$h$ = mappedHost($j$) \\
\keyw{if} ( \( \exists ar_1, \ldots, ar_m  \in ARESOURCE, \exists pr_1, \ldots, pr_m  \in PRESOURCE \) ): \\
\> request($j$, $ar_i$) \( = true  ~ \& ~ h \neq undef \) \\
\> \( \& ~ \) canUse(user($j$), $pr_k$) \( = true \), belongsTo($pr_k$, $h$) \( = true, 1 \leq i, k \leq m \) \\
\> \> \keyw{then} submitted($j$, $h$) \( := true \) \\
\keyw{endif} \\
\end{tabbing}

\noindent \(  \Pi_{\mathrm{host\_mapping}} \) \\

\begin{tabbing}
\quad \=\quad \=\quad \=\quad \=\quad \kill
\keyw{if} ( \( \exists j \in JOB, \exists ar_1, \ldots, ar_m  \in ARESOURCE, \exists pr_1, \ldots, pr_m  \in PRESOURCE \) ): \\
\> mappedHost($j$) \( = undef ~ \& \) request($j$, $ar_i$) \( = true, 1 \leq i \leq m \) \\
\> \> \keyw{then} \keyw{choose} $h$ \keyw{in} $HOST$ \\
\> \> \> \keyw{satisfying} compatible(attr($ar_i$), attr($pr_k$)) \\
\> \> \> \keyw{where} belongsTo($pr_k$, $h$) \( = true, 1 \leq i, k \leq m \) \\
\> \> \> \> mappedhost($j$) \( := h \) \\
\> \> \keyw{endchoose} \\
\keyw{endif} \\
\end{tabbing}

\subsection{Rule 5: Broker selection}

At the highest level of Grid resource management a broker needs to be selected automatically for a user job. An important precondition 
of the selection process is that such a broker needs to be selected that manages hosts with resources that the user of the job can use. 
Furthermore the agent responsible for broker selection, the meta-broker (universe $METABROKER$) needs to ensure that the chosen broker 
has all the properties required by the user's job. Therefore users need to characterize their job requirements in a certain job description 
language, which should include both the required broker properties and abstract resources of the processes of the job. This additional 
Grid middleware component is highlighted in Figure \ref{fig3} (b). The following function is added to the model: \\ 

\noindent mappedBroker: \( JOB \rightarrow BROKER \) \\

\noindent We extend the initial state by: \\

\noindent \( \forall j: \) mappedBroker($j$) \( = undef \) \\

\noindent The formal definition of the meta-broker agent is as follows: \\

\begin{tabbing}
\quad \=\quad \=\quad \kill
\keyw{let} \( b = \) mappedBroker($j$) \\
\keyw{if} ( \( \exists r \in REQUIREMENT, \exists pr \in PRESOURCE, \exists h \in HOST \) ):  \\
\> request($j$, $r$) \( = true ~ \& ~ b \neq undef ~ \& \) canUse(user($j$), $pr$) \( = true \), \\
\> belongsTo($pr$, $h$) \( = true \), manages($h$, $b$) \( = true \) \\
\> \> \keyw{then} submitted($j$, $b$) \( : = true \) \\
\keyw{endif} \\
\end{tabbing}

\noindent \(  \Pi_{\mathrm{broker\_mapping}} \) \\

\begin{tabbing}
\quad \=\quad \=\quad \=\quad \kill
\keyw{if} ( \( \exists r_1, \ldots, r_m  \in REQUIREMENT, \exists p_1, \ldots, p_m  \in PROPERTY, \exists j \in JOB, \) \\
\( \exists b \in BROKER \) ):  \\
\> mappedBroker($j$) \( = undef ~ \& ~ \forall i: \) request($j$, $r_i$) \( = true \) \\
\> \( ~ \& ~ \forall i: \) have($b$, $p_i$) \( = true, ~ 1 \leq i \leq m \) \\
\> \> \keyw{then} \keyw{choose} $b$ \keyw{in} $BROKER$  \\
\> \> \> \keyw{satisfying} compatible(attr($r_i$), attr($p_i$)), \( 1 \leq i \leq m \)  \\
\keyw{endif} \\
\end{tabbing}

Finally we should state that jobs can be interconnected in order to form a complex Grid application called as workflows. 
The execution of workflows require a coordinating tool called workflow enactor that schedules the interdependent 
jobs for executions. We refrain from formalizing workflow management and incorporate it into our model, since our central 
entities are jobs, and therefore assume that grid applications are submitted into the system in the form of jobs. 

As a summary, we have shown that grid brokering takes place at three levels, and the following 
operations need to be performed: broker mapping, host mapping and resource mapping. In the following section we show, how 
practical examples of these components can be described by our formal ASM model with the help of ASM refinement. These tools 
are the Grid Meta-Broker \cite{mb} and GTbroker \cite{gridintop}.

\section{Refinement of Broker Components}\label{refinement}

This section contains illustrative examples, how the generic brokering model can be refined into models that represent realised implementations 
of the brokering principles. One can see in these examples how certain functions, kept abstract in Rule 4 and 5 presented earlier, are transformed 
to reveal implementation details. More information on the realization and practical utilization of these tools can be read in \cite{gridintop}.

\subsection{Refinement of broker mapping (matchmaking of the Grid Meta-Broker)}

\noindent \(  \Pi ' _{\mathrm{broker\_mapping}} \) \\

\begin{tabbing}
\quad \=\quad \=\quad \=\quad \=\quad \kill
\keyw{if} ( \( \exists r_1, \ldots, r_m \in REQUIREMENT, \exists p_1, \ldots, p_n \in PROPERTY, \) \\
\( \exists j \in JOB, \exists b_1, \ldots, b_l \in BROKER, \exists v_1, \ldots, v_l \in REAL \) ): \\
\> mappedbroker($j$) \( = undef ~ \& ~ \forall t: \) request($j$, $r_t$) \( = true, 1 \leq t \leq m, \) \\
\> \( \& ~ \forall i: \) have($b_k$, $p_i$) \( = true, 1 \leq i \leq n , 1 \leq k \leq l  \) \\
\> \> \keyw{then do forall} \( k ~ (1 \leq k \leq l) \) \\
\> \> \> $v_k$ $:=$ getBrokerPerf($b_k$) \\
\> \> \> \keyw{if} ( \( \neg \exists t, i \) ): attr($r_t$) $=$ attr($p_i$) $\&$ have($b_k$, $p_i$) \( = true, 1 \leq t \leq m, 1 \leq i \leq n  \) \\
\> \> \> \> \keyw{then} $v_k$ $:=$ $0$ \\
\> \> \keyw{enddo} \\
\> \> \keyw{choose} $v_{max}$ \keyw{in} ( \( v_1, \ldots, v_l \) ) \\
\> \> \> \keyw{satisfying} \( v_{max} \geq v_k, 1 \leq k, max \leq l \) \\
\> \> \> \> mappedbroker($j$) $:=$ $b_{max}$ \\
\> \> \keyw{endchoose} \\
\keyw{endif} \\
\end{tabbing}

In addition to the broker mapping defined in Rule 1, this refinement details how the \emph{compatible} function 
is implemented. In case of the Grid Meta-broker, the attributes of the broker properties are certain keywords. 
The users have to use the same keywords in their requirement specifications, therefore compatibility means 
exact string matching. The refined agent also uses an additional function getBrokerPerf: \( BROKER \rightarrow REAL \), 
which returns a real number denoting the dynamic performance of the appropriate broker. The higher this value is the 
better the broker performs. 

\subsection{Refinement of host mapping (matchmaking of GTbroker)}

\noindent \(  \Pi ' _{\mathrm{host\_mapping}} \) \\

\begin{tabbing}
\quad \=\quad \=\quad \=\quad \=\quad \kill
\keyw{if} ( \( \exists j \in JOB, \exists ar_1, \ldots, ar_n \in ARESOURCE, \exists policy \in REQUIREMENT, \) \\
\( \exists pr_1, \ldots, pr_m \in PRESOURCE, \exists h_1, \ldots, h_t \in HOST, \exists r_1, \ldots, r_t \in REAL \) ): \\
\> mappedhost($j$) \( = undef ~ \& ~ \) request($j$, $policy$) \( = true, \) \\
\> request($j$, $ar_i$) \( = true, 1 \leq i \leq n \leq m \) \\
\> \> \keyw{then do forall} \( k ~ (1 \leq k \leq t) \) \\
\> \> \> $r_k$ $:=$ countRank($policy$, $h_k$) \\
\> \> \> \keyw{if} ( \( \neg \exists l, i \) ): attr($ar_i$) $\leq$ attr($pr_i$) \\
\> \> \> $\&$ belongsTo($pr_l$, $h_k$) \( = true, 1 \leq i \leq n, 1 \leq l \leq m  \) \\
\> \> \> \> \keyw{then} $r_k$ $:=$ $0$ \\
\> \> \keyw{enddo} \\
\> \> \keyw{choose} $r_{max}$ \keyw{in} ( \( r_1, \ldots, r_t \) ) \\
\> \> \> \keyw{satisfying} \( r_{max} \geq r_k, 1 \leq k, max \leq t \) \\
\> \> \> \> mappedhost($j$) $:=$ $h_{max}$ \\
\> \> \keyw{endchoose} \\
\keyw{endif} \\
\end{tabbing}

In addition to the host mapping defined in Rule 2, this refinement also reveals the meaning of the \emph{compatible} 
function. In case of GTbroker, the attributes of resource requirements denote the amount of resource capacity (e.g. 
memory size or processor speed) needed by the processes of the job for execution. This means, if the available physical 
resource has equal or greater capacity than requested, the process can run. The host selection method can be influenced 
by users using the special $policy$ requirement. The value of its attribute tells the additional countRank: \( REQUIREMENT \times HOST \rightarrow REAL \) 
function how to compute the rank for the available hosts (e.g. higher priority can be given to hosts with faster processors). 
Finally, the host with the highest rank is selected for execution.

\section{Conclusions}\label{conclusions}

In this paper we have investigated the brokering components of the Grid middleware and defined a three-layered 
\emph{formal model} using \emph{Abstract State Machines}. In this model three agents are responsible for resource management by 
performing three selection processes: broker mapping, host mapping and resource mapping. We have also proposed \emph{two refined 
definitions} for broker and host selection, which are implemented by the Grid Meta-broker and GTbroker. Our future work 
aims at introducing interoperability metrics for categorizing brokering components and using the ASM model for verifying our 
categorization.

\bibliographystyle{eptcs}

\end{document}